\documentclass[prd,amsmath,amssymb,nofootinbib,superscriptaddress,twocolumn,10pt]{revtex4-1}

\pdfoutput=1

\usepackage{CJKutf8}
\usepackage{graphicx}
\usepackage{amsmath}
\usepackage{subfigure}
\usepackage{xcolor}
\definecolor{BLUE}{named}{blue}
\usepackage{siunitx}
\usepackage{float}
\usepackage[colorlinks=true, linkcolor=red, citecolor=blue]{hyperref}

\begin{document}

\title{Gaussian curvature and Lyapunov exponent as probes of black hole phase transitions}

\author{Shi-Hao Zhang}
\affiliation{Liaoning Key Laboratory of Cosmology and Astrophysics, College of Sciences, Northeastern University, Shenyang 110819, China}
\author{Zi-Qiang Zhao}
\affiliation{Liaoning Key Laboratory of Cosmology and Astrophysics, College of Sciences, Northeastern University, Shenyang 110819, China}
\author{Zi-Yuan Li}
\affiliation{School of Physics,
    Nankai University, Tianjin 300071, China}
\author{Jing-Fei Zhang}
\email{jfzhang@mail.neu.edu.cn}
\affiliation{Liaoning Key Laboratory of Cosmology and Astrophysics, College of Sciences, Northeastern University, Shenyang 110819, China}
\author{Xin Zhang}
\email{zhangxin@mail.neu.edu.cn}
\affiliation{Liaoning Key Laboratory of Cosmology and Astrophysics, College of Sciences, Northeastern University, Shenyang 110819, China}
\affiliation{Key Laboratory of Data Analytics and Optimization for Smart Industry (Ministry of Education), 
	Northeastern University, Shenyang 110819, China}
\affiliation{National Frontiers Science Center for Industrial Intelligence and Systems Optimization, 
Northeastern University, Shenyang 110819, China}


\begin{abstract} 

First-order phase transitions of black holes have been extensively studied within thermodynamic frameworks, yet the corresponding evolution of spacetime geometric properties remains unclear. This paper establishes a purely differential geometric framework to probe such phase transitions by analyzing the curvature of unstable null orbits. Using the geodesic curvature of the null circular orbit in the optical metric to locate the light ring, we demonstrate that the corresponding Gaussian curvature $K$ serves as a direct geometric signature of the phase transition. During a first-order phase transition, the curve $K$ versus temperature $T$ exhibits a multivalued structure within the spinodal region, precisely mirroring the swallowtail behavior of the free energy. Numerical analysis of Hayward-Letelier-AdS black holes confirms the effectiveness of this geometric signature. Our work demonstrates that the intrinsic geometric quantities of spacetime encode the information of black hole phase transitions. These quantities serve as geometric probes of black hole phase transitions, while their discontinuity between the small and large black hole branches exhibits order parameter-like behavior. As an extension of this geometric probe, we also find that the Gaussian curvature exhibits a heat‑capacity‑like divergence at the second‑order phase transition point. These results provide a purely geometric foundation for understanding the correspondence between thermodynamics and spacetime curvature in the null case.

\end{abstract}
	

\maketitle

\section{Introduction}\label{chap:1}

Since the 1970s studies by Bekenstein and Hawking \cite{Hawking:1975vcx,Bekenstein:1973ur}, black holes have been regarded as thermodynamic systems. Subsequent studies revealed that black holes can undergo phase transitions, such as the Hawking-Page phase transition \cite{HPP}.  Later studies showed that when the cosmological constant is treated as pressure, black holes exhibit a van der Waals-like first-order phase transition \cite{P} and a reentrant phase transition \cite{REP}. 
Furthermore, building on the holographic principle in black hole thermodynamics, Maldacena established the AdS/CFT correspondence \cite{Maldacena:1997re} in 1997. Further investigation within this duality showed that black holes as quantum thermodynamic systems exhibit chaos and that their Lyapunov exponents obey the Maldacena-Shenker-Stanford (MSS) bound $\lambda \leq 2\pi T/\hbar$ \cite{MSS01,MSS02,MSS03}. In recent years, some studies have applied classical dynamics methods to verify the validity of the MSS bound \cite{TMSS01,TMSS02,TMSS03,TMSS04,TMSS05,Lee:2025ias}. The Lyapunov exponent not only directly characterizes the chaotic behavior of black holes as quantum thermodynamic systems, but has also been connected to the imaginary part of quasinormal modes (QNMs) \cite{QNM01,QNM02}.

In black hole thermodynamics, the multivaluedness of QNMs \cite{QNMP01,QNMP02,QNMP03,QNMP04,QNMP05,Zhao:2022jvs,Zhao:2023ffs,Zhao:2024jhs,QNMP06,Hou:2025bli}  and Lyapunov exponents \cite{JHEP2022,BIADS,GB,Kumara:2024obd,KR,order,HaywardADS,KADS,EPYM,Rcharge,KNADS,Cheng:2026dnd,Lee:2026qtn} during first-order phase transitions suggest a potential connection between these quantities and black hole phase transitions. Studies of the observable photon sphere have also tied it to black hole phase transitions \cite{PS01,PS02,PS03,PS04,PS05,PS06,PS07,PS08,PS09}, confirming that dynamical analysis can probe such phenomena. However, one question remains largely unexplored. Since general relativity is fundamentally geometric, how does spacetime geometry itself change during a thermal phase transition? Is there an intrinsic geometric quantity that can directly reflect this phase transition?

Black holes are solutions of the Einstein equations whose physical processes are reflected by changes in geometric properties. Gaussian curvature, an intrinsic measure of a two-dimensional manifold, directly quantifies local spacetime curvature and serves as an ideal probe of geometry. Using Gaussian curvature, researchers have developed purely geometric methods to study black hole photon spheres \cite{GC01,CWH}, thus establishing a clear correspondence between dynamics and geometry. Recently, it has been used to analyze the stability of circular orbits for massless particles, to determine the existence and distribution of stable and unstable orbits \cite{GC0X}, and to investigate gravitational lensing by massive particles \cite{GCGL01,GCGL02}. This raises a previously overlooked question: can geometric quantities such as Gaussian curvature be used to connect with first-order phase transitions of a black hole? A key finding is that the Lyapunov exponent of null circular orbits near black holes is linked to their Gaussian curvature \cite{L=K}. The established role of the Lyapunov exponent as an effective probe makes it possible to address the above question.

In this work, we investigate unstable particle orbits near a 3+1-dimensional spherically symmetric black hole. We demonstrate that the Gaussian curvature of unstable orbits of massless particles is multivalued at the black hole first-order phase transition point, offering a new perspective for understanding black hole phase transitions. This reveals that during such phase transitions, characterized by the existence of multiple spacetime solutions, the corresponding geometric quantity exhibits multivalued behavior. The Appendix also shows that the divergence of the derivative of the Gaussian curvature at the second‑order phase transition point offers an intrinsic geometric criterion for second‑order phase transitions.

This paper is arranged as follows. In Sec.~\ref{chap:2}, we briefly review the relations of black hole thermodynamics, the Gaussian curvature of two-dimensional surfaces, and the calculation of Lyapunov exponents. In Sect.~\ref{chap:3}, we present the investigation of the Hayward-Letelier-AdS black hole and the analysis of the results. In Sec.~\ref{chap:4}, we address the computation of scaling exponents. In Sec.~\ref{chap:5}, we offer a summary and discussion, and the appendix follows. We set $G=c=k_B=\hbar=1$ in this paper.

\section{Theoretical Framework} \label{chap:2}

In this section, we first briefly review the relations of black hole thermodynamics, then introduce how to obtain the Gaussian curvature from a two-dimensional Riemannian metric, and finally recall the derivation of the Lyapunov exponent and its connection to the Gaussian curvature.

\subsection{Thermodynamics and phase structure of spherically symmetric black holes}
In this subsection, we briefly review the relations in black hole thermodynamics by considering a 3+1-dimensional spherically symmetric black hole solution
\begin{align}
   ds^{2}=-f(r)dt^{2}+\frac{1}{g(r)}dr^{2}+r^2 \left( d\theta^2 + \sin^2\theta \, d\phi^2 \right)\label{metric 01},
\end{align}
where $f(r)$ and $g(r)$ are smooth functions of class $C^P$ ($P\geq2$).
Note that $f(r_+)=0$, where $r_+$ is the horizon radius. The temperature is given by
\begin{align}
T = \frac{\sqrt{f'g'}}{4\pi} \bigg|_{r_+},\label{T}
\end{align}
for a fixed charge and pressure. The Gibbs free energy is
\begin{align} 
F = \mathcal{M}-TS, \label{F}
\end{align} 
where $S =A/4 =\pi r_+^2$ is the entropy, $A$ is the area of the black hole horizon, $\mathcal{M}$ is the corrected mass (if no correction is needed, it is the ADM mass of the black hole).
If a first-order phase transition occurs, the free energy exhibits a swallowtail structure, indicating the appearance of three black hole solutions (large black hole, intermediate black hole, and small black hole). The critical point is determined by the following condition
\begin{align}
\frac{\partial T}{\partial r_+} = \frac{\partial^2 T}{\partial r_+^2} = 0.
\end{align}
For an Anti-de Sitter spacetime, the pressure is given by \cite{P}:
\begin{align} 
P = -\frac{\Lambda}{8\pi} = \frac{3}{8\pi \ell^2},
\end{align}
where $\Lambda$ is a cosmological constant and $\ell$ is the AdS radius.

\subsection{Gaussian curvature of two-dimensional manifolds}
Gaussian curvature $K$ is an intrinsic geometric quantity of the optical metric associated with massless particles, and the light ring is closely linked to black hole spacetime. This suggests that $K$ on the light ring may exhibit anomalous behavior near the phase transition point, offering a potential geometric signature for studying such phase transitions.

To analyze null geodesics, we rewrite Eq.~(\ref{metric 01}) as the optical metric ($ds^{2}=0$), and restrict our attention to the metric in the equatorial plane ($\theta= \frac{\pi}{2}$)
\begin{align}
dt^2 = \frac{1}{f(r)} \left( \frac{1}{g(r)} dr^2 + r^2 d\phi^2 \right).\label{metric 02}
\end{align}
Beginning with Eq.~(\ref{metric 02}), which describes a two-dimensional Riemannian manifold, we compute the Gaussian curvature associated with distinct null circular orbits. For a two-dimensional Riemannian manifold in orthogonal coordinates, its Gaussian curvature is given by
\begin{align}
K=-\frac{1}{2\sqrt{EG}} \left\{ \frac{\partial}{\partial v} \left[ \frac{(E)_v}{\sqrt{EG}} \right] + \frac{\partial}{\partial u} \left[ \frac{(G)_u}{\sqrt{EG}} \right] \right\},\label{KG}
\end{align}
where $(u,v)$ are the coordinate variables on the surface, and $E, G$ are the coefficients of the first fundamental form \cite{CWH}. By setting $E=g_{11}$ and $G=g_{22}$, Eq.~(\ref{KG}) translates Gaussian notation into tensor notation
\begin{align}
K = -\frac{1}{2} \frac{1}{\sqrt{g_{rr} g_{\phi\phi}}} \frac{d}{dr} \left( \frac{g'_{\phi\phi}}{\sqrt{g_{rr} g_{\phi\phi}}} \right)\label{Gaussian curvature}.
\end{align}
Substituting Eq.~(\ref{metric 02}) into (\ref{Gaussian curvature}) yields
\begin{equation}
\begin{split}
K(r) =& -g'(r) \frac{2f(r) - rf'(r)}{4r} \\
     & + \frac{g(r)}{2} \left[ f'(r) \left( \frac{1}{r} - \frac{f'(r)}{f(r)} \right) + f''(r) \right]
\label{Gaussian curvature 01}.
\end{split}
\end{equation}
For a circular orbit in the equatorial plane of the optical metric Eq.~(\ref{metric 02}), its geodesic curvature $\kappa_g$ is given by
\begin{align}
\kappa_g = \sqrt{\frac{g(r)}{f(r)}} \frac{2f(r) - rf'(r)}{2r}.
\end{align}
On the unstable null circular orbit $r=r_{LR}$ (the light ring), the geodesic curvature vanishes, yielding
\begin{align}
2f(r_{LR}) = r_{LR} f'(r_{LR}).\label{Null 01}
\end{align}
For the unstable null circular orbit ($r=r_{LR}$, inserting Eq.~(\ref{Null 01}) into (\ref{Gaussian curvature 01}) yields a known relation \cite{L=K}
\begin{align}
K(r_{LR}) = \left\{ \frac{g(r)}{2} \left[ f''(r) - \frac{f'(r)}{r} \right] \right\} \bigg|_{r=r_{LR}}.\label{Gaussian curvature 02}
\end{align}
Note that $K(r_{LR})$ depends only on the derivatives of the spherically symmetric black hole metric functions $g(r)$ and $f(r)$. 
As an intrinsic geometric quantity of a surface, the Gaussian curvature depends solely on the first fundamental form of the surface (a profound result known as the \textit{Egregium Theorem}). It quantifies the deviation of the first fundamental form of the two-dimensional surface from the Euclidean metric, which is fundamental to the study of intrinsic geometry \cite{CWH}. As a coordinate-invariant intrinsic curvature, $K$ directly measures spacetime deformation, unlike orbital stability analyses.

In Sec.~\ref{chap:3}, we will demonstrate the connection between Gaussian curvature and first-order phase transitions by examining the anomalous behavior of $K$ for circular null orbits near the Hayward-Letelier-AdS. A more explicit link between the detailed behavior of $K$ and the phase transition will be provided in the next subsection through the analysis of Lyapunov exponents.

\subsection{Lyapunov exponent for unstable orbits}
The Lyapunov exponent $\lambda$ is a dynamical quantity that characterizes the chaotic behavior of particle orbits near black holes, and it does not directly reflect a phase transition. In this subsection, we briefly review the derivation of $\lambda$  and its relationship with $K$.

Beginning with the metric Eq.~(\ref{metric 01}), the Lagrangian is
\begin{align}
\mathcal{L} = g_{\mu\nu} \dot{x}^\mu \dot{x}^\nu = \sigma,\label{Lagrangian}
\end{align}
where $\sigma = \{+1,-1,0\}$ denotes spacelike, timelike, and null geodesics, respectively.
The angular momentum and energy are defined by
\begin{align}
2L = 2 \frac{\partial \mathcal{L}}{\partial \dot{\phi}} = 2r^2 \dot{\phi}, \quad -2E = 2 \frac{\partial \mathcal{L}}{\partial \dot{t}} = -2f \dot{t}.\label{LE}
\end{align}
On the equatorial plane, the radial motion follows
\begin{align}
(\dot{r})^2 = V_{\text{eff}}(r),
\end{align}
with effective potential
\begin{align}
V_{\text{eff}} (r)= g(r) \left[ \frac{E^2}{f(r)} - \frac{L^2}{r^2} - \sigma \right].
\end{align}
For massless particles $\sigma=0$, the second derivative on light rings $r_{LR}$ satisfies
\begin{align}
 V''_{\text{eff}}(r_{LR}) = \frac{g(r) L^2}{r^4 f(r)} \left[ r f'(r) - r^2 f''(r) \right] \bigg|_{r=r_{LR}}.
\end{align}
Expanding this equation of motion at the light ring yields
\begin{align}
(\dot{\varepsilon})^2 - \frac{1}{2} V''_{\text{eff}}(r_{LR}) \varepsilon^2 = 0,
\end{align}
where $\varepsilon=r-r_{LR}$. Note that this is an inverted harmonic oscillator equation, defining the Lyapunov exponent \cite{EOML,QNM01}
\begin{align}
\lambda = \sqrt{\frac{1}{2(\dot{t})^2} V''_{\text{eff}}(r_{LR})}.\label{lambda}
\end{align}
The general solution is
\begin{align}
\varepsilon = A e^{\lambda t} + B e^{-\lambda t}.
\end{align}
Combining Eqs.~(\ref{LE}),~(\ref{lambda}), and~(\ref{Gaussian curvature 02}) yields a known relation \cite{L=K}
\begin{align}
K(r_{LR}) = -\lambda^2(r_{LR}).\label{lambdaK}
\end{align}
Eq.~(\ref{lambdaK}) establishes a quantitative relation between Gaussian curvature and Lyapunov exponent on the light ring. Previous studies \cite{JHEP2022,BIADS,GB,KR,order,HaywardADS,KADS,EPYM,Rcharge,KNADS,Cheng:2026dnd,Lee:2026qtn} have demonstrated that $\lambda$ becomes multivalued near the phase transition point. In Refs.~\cite{JHEP2022,order}, the authors interpreted the Lyapunov exponent as an order parameter that decreases with increasing temperature $T$. Clearly, Eq.~(\ref{lambdaK}) provides a mapping between $K$ and $\lambda$ on the light ring. Such a mapping paves the way for a connection between geometry and thermodynamics, thereby enabling the investigation of thermodynamics from a geometric perspective. Combined with Eq.~(\ref{lambdaK}), if $\lambda$ is multivalued in the spinodal region, $K$ must also be multivalued, offering a geometric way to probe black hole phase transitions. Note that not only do null circular orbits admit a Lyapunov exponent, but circular orbits of massive particles can also be defined. In both cases, these dynamical quantities connect with the phase transition.

For completeness, we also examine the Lyapunov exponent associated with the circular orbits of massive particles. The corresponding unstable circular orbit is located at $r=r_0$ with $\sigma=-1$, and its Lyapunov exponent is given by
\begin{equation}
\begin{split}
\lambda &= \frac{1}{2} \left. \sqrt{[2f(r)-rf'(r)]V_{\rm eff}''(r)} \right|_{r=r_0} \\
&= \frac{1}{\sqrt{2}} \sqrt{-\frac{g(r_0)}{f(r_0)} \left[ \frac{3f(r_0)f'(r_0)}{r_0}-2f'(r_0)^2+f(r_0)f''(r_0) \right]}.
\end{split}
\end{equation}
For $g(r)=f(r)$, the conserved quantities satisfy
\begin{align}
\frac{E}{L} = \frac{\sqrt{f(r_0)}}{r_0}.
\end{align}

In Sec.~\ref{chap:3}, we also examine the relationship between $\lambda$ and $\tilde{T}$, where the tilde denotes dimensionless quantities, for massless and massive particles around a specific black hole.

\section{Numerical Results and Analysis}\label{chap:3}
In Sec.~\ref{chap:2}, we established a theoretical framework valid for static 3+1-dimensional spherically symmetric black holes. This section probes the Hayward–Letelier–AdS black hole  (with $f(r)=g(r)$), which avoids spacetime singularities and possesses a nontrivial matter source, thereby offering a nontrivial geometric spacetime model beyond the singular electrovacuum case, and we examine our conjecture by following these steps. We first examine the $\tilde{F}(\tilde{T})$ curve to determine the spinodal region (the temperature interval where small/intermediate/large phases coexist) $(\tilde{T}_1, \tilde{T}_2)$ from its characteristic multivaluedness. We then investigate whether the corresponding $K(\tilde{T})$ and $\lambda(\tilde{T})$ curves also become multivalued within the same temperature interval.

We investigate the behavior of $\lambda(\tilde{T})$ and $K(\tilde{T})$ for null geodesics near the phase transition point $\tilde{T}_p$ in the Hayward–Letelier–AdS black hole. This solution corresponds to a Hayward black hole in AdS spacetime surrounded by a cloud of strings \cite{HLT01}. Its metric function is
\begin{align}
f_{HL}(r) = 1 - \frac{2Mr^2}{g^3 + r^3} + \frac{r^2}{\ell^2}-a,\label{fhl}
\end{align}
where $g$ is the magnetic monopole charge, $\ell$ is the AdS radius, and the string-cloud parameter $a$ arises from the string-cloud term in action \cite{HLT01}. For studies of Hayward-AdS black holes with string fluids in the extended thermodynamic phase space, see Ref.~\cite{Anand:2025mlc} for thermodynamic curvature and topological properties, and Ref.~\cite{Singh:2025kaz} for chaotic behavior. When $a=0$, the metric reduces to the Hayward–AdS black hole. The temperature is \cite{HLT01}
\begin{align}
T_{HL} = \frac{r_h^3\bigl[(1-a) + 3r_h^2/\ell^2\bigr] - 2g^3(1-a)}{4\pi r_h (r_h^3 + g^3)}.
\end{align}
After considering the correction to the mass, a self-consistent free energy is obtained
\begin{widetext}
\begin{align}
\begin{split}
F &= \frac{1}{2\ell^2} \Bigg[ -(a-1)\ell^2 r_h + r_h^3 + \sqrt{3}(a-1)g\ell^2 \arctan\!\left(\frac{2r_h-g}{\sqrt{3}g}\right) \\
&\qquad + (a-1)g\ell^2 \ln(g+r_h) - \frac{1}{2}(a-1)g\ell^2 \ln(g^2 - g r_h + r_h^2) - g^3 \ln(g^3 + r_h^3) \Bigg] \\
&\qquad - \frac{r_h^3\bigl((1-a) + 3r_h^2/\ell^2\bigr) - 2g^3(1-a)}{4\pi r_h (r_h^3 + g^3)} \, \pi r_h^2.
\end{split}
\label{FHL}
\end{align}
\end{widetext}
Here,
\begin{align}
M = \left(1 + \frac{r_+^2}{\ell^2}-a\right) \frac{g^3 + r_+^3}{2r_+^2}.
\end{align}
Since $a$ is a dimensionless constant, We introduce the following scaling
\begin{align}
\tilde{r}_+ = \frac{r_+}{\ell}, \quad \tilde{g} = \frac{g}{\ell}, \quad \tilde{M} = \frac{M}{\ell}, \quad \tilde{F}_H = \frac{F_H}{\ell}, \quad \tilde{T}_{H} = T_{H}\ell\label{scaling}.
\end{align}
We present the $\tilde{F}_{HL}(\tilde{T}_{HL})$, $K_{HL}(\tilde{T}_{HL})$, and $\lambda_{HL}(\tilde{T}_{HL})$ curves for the Hayward–Letelier–AdS black hole.

\begin{figure*}
	\begin{minipage}{1\hsize}
		\begin{center}
			
            \subfigure[$\tilde{F}_{HL}-\tilde{T}_{HL}$]{
				\label{HLFT}
				\includegraphics*[scale=0.27]{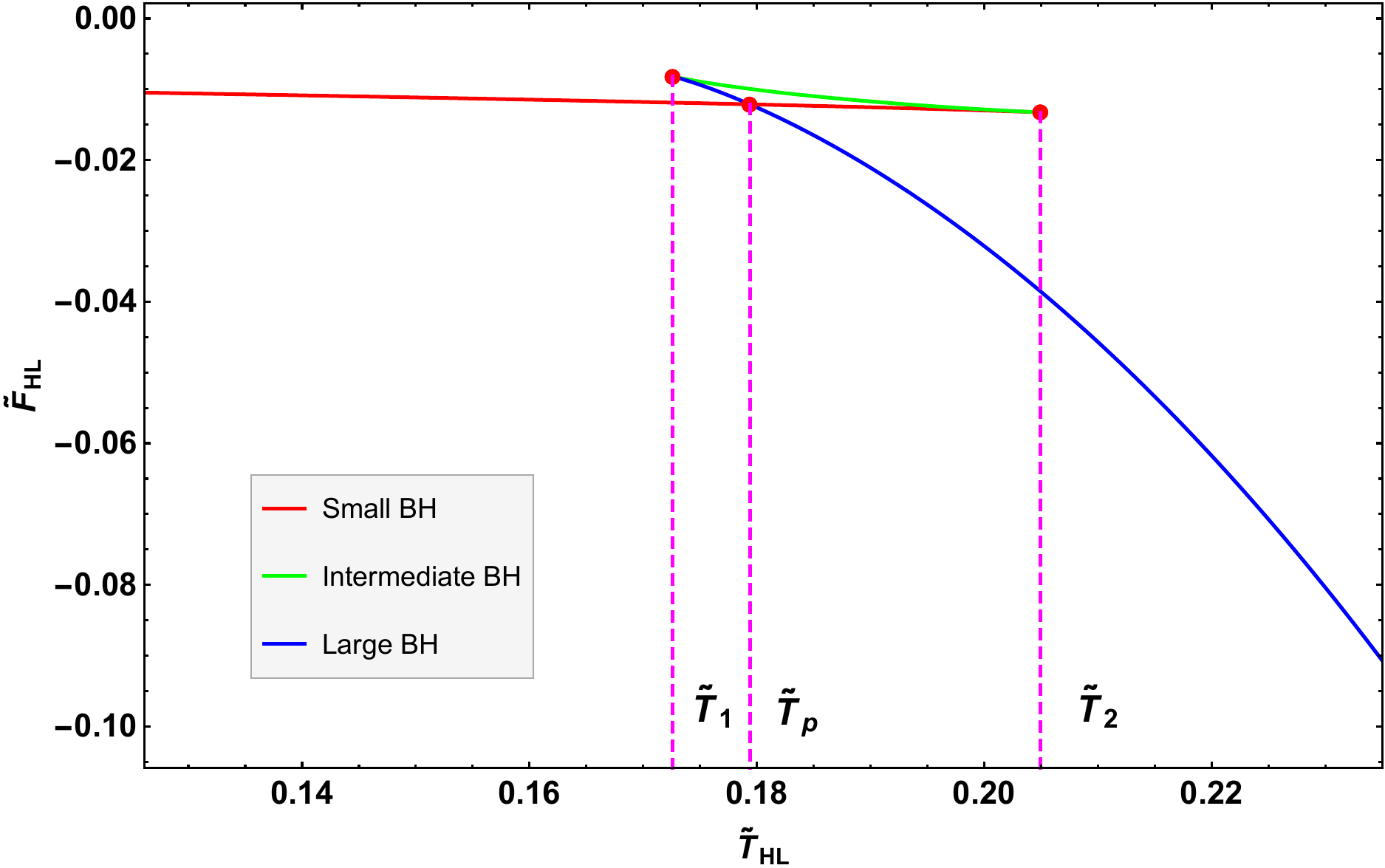}}
            \subfigure[$K_{HL}-\tilde{T}_{HL}$]{
				\label{HLKT}
				\includegraphics*[scale=0.27]{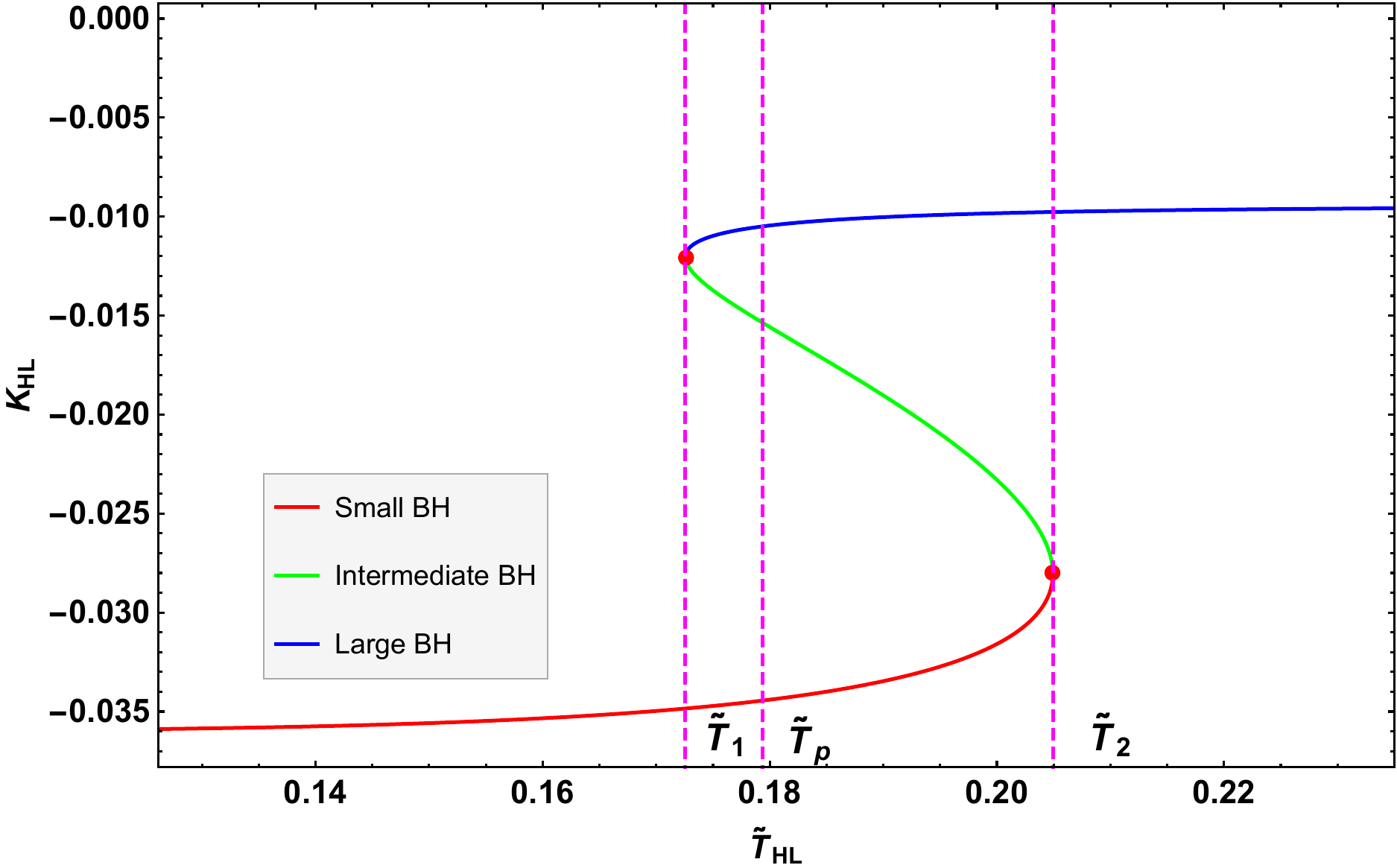}}    
			\subfigure[$\lambda_{HL}-\tilde{T}_{HL}$ (Null)]{
				\label{HLNullLT}
				\includegraphics*[scale=0.27]{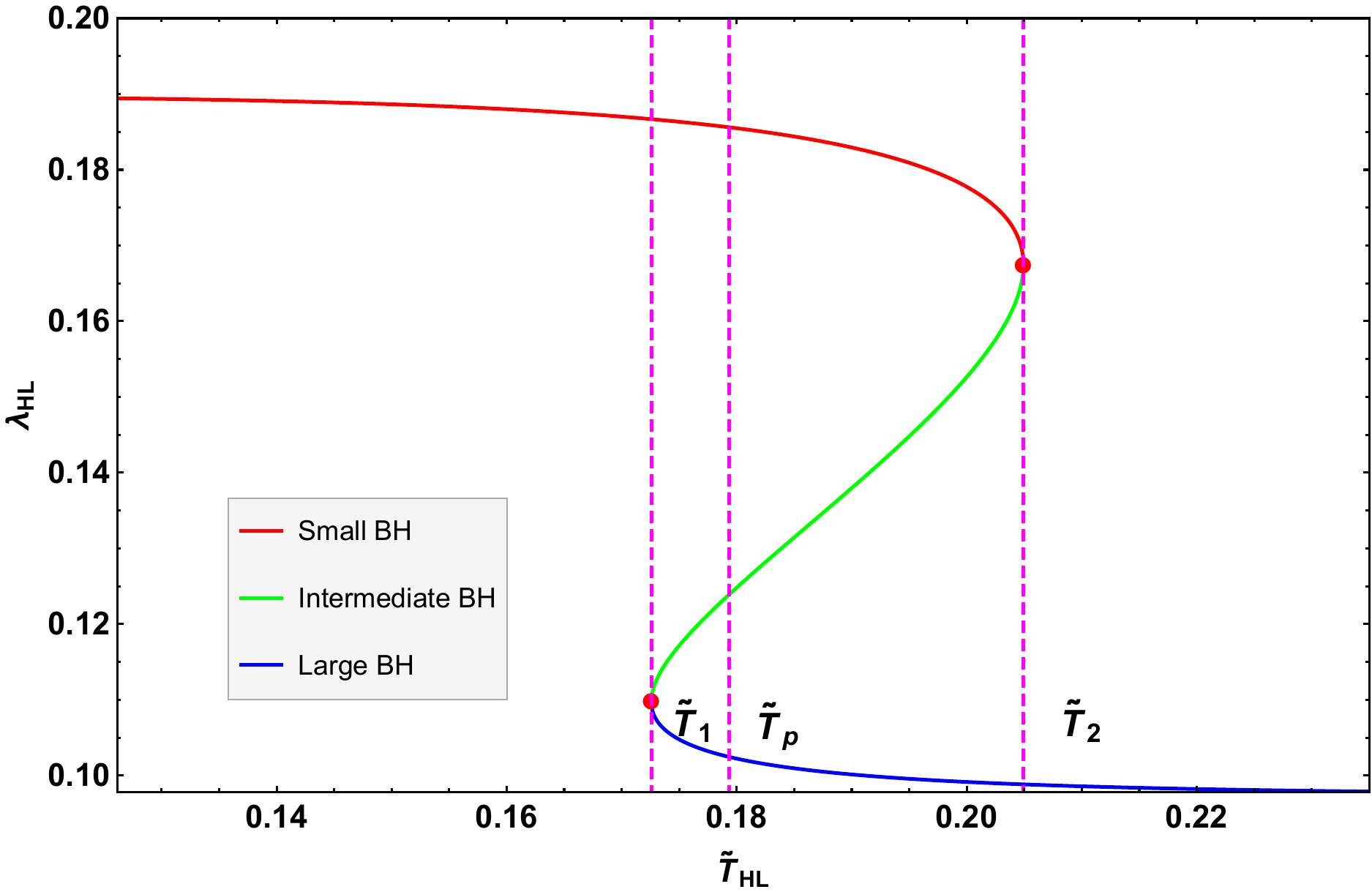}}
			\subfigure[$\lambda_{HL}-\tilde{T}_{HL}$ (Timelike)]{
				\label{HLTimelikeLT}
				\includegraphics*[scale=0.256]{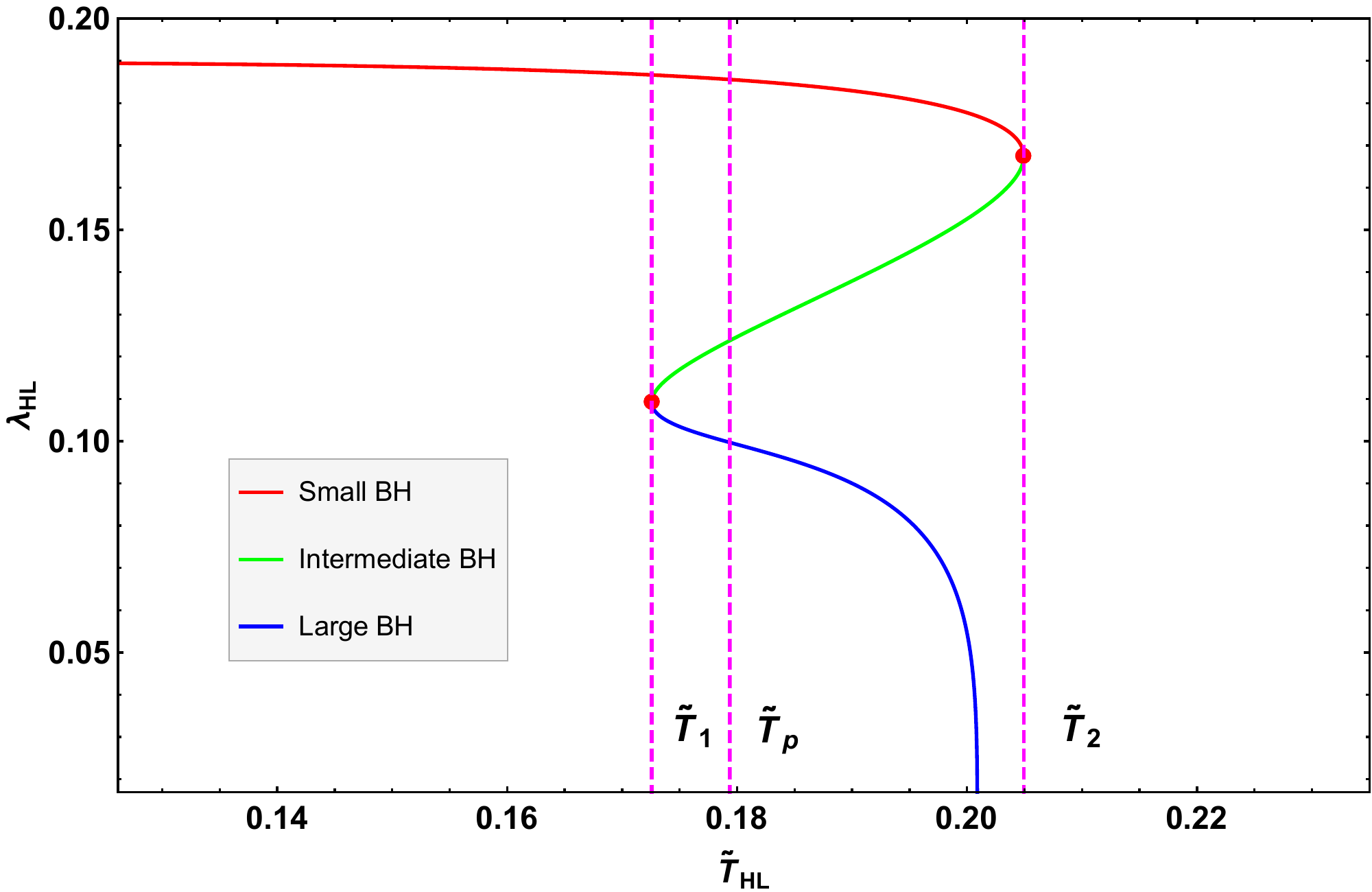}}
		\end{center}
		\caption{Thermodynamic, geometric and chaos signatures of the first-order phase transition in Hayward-Letelier-AdS space-times. (a) $\tilde{F}_{HL}$, (b) $K_{HL}$ of unstable null orbits and (c-d) $\lambda_{HL}$ of unstable null/timelike orbits versus temperature $\tilde{T}_{HL}$ for Hayward-Letelier-AdS black holes, and $\tilde{g}=0.061,~\tilde{g}_c=0.090,~a=0.6,~\tilde{g}<\tilde{g}_c,~L=20\ell$. Swallowtail structures in $\tilde{F}_{HL}$ and multivalued $K_{HL}$ at $\tilde{T}_{p}$ evidence geometric degeneracy during phase transitions. $\lambda_{HL}$ at $\tilde{T}_{p}$ also exhibits such multivaluedness.
        }
		\label{Fig.3}
	\end{minipage}
\end{figure*}

As illustrated in Fig.~\ref{HLFT}, for $\tilde{g}<\tilde{g}_c$ the swallowtail structure indicates a first-order phase transition. Similarly, in Figs.~\ref{HLKT}, \ref{HLNullLT}, and \ref{HLTimelikeLT}, the $K_{HL}(\tilde{T}_{HL})$ curve for null circular orbits and the $\lambda_{HL}(\tilde{T}_{HL})$ curve for both timelike and null orbits exhibit multivalued behavior. This shows that both the Gaussian curvature $K_{HL}$ and the Lyapunov exponent $\lambda_{HL}$ are intimately related to the first-order phase transition, because Eq.~(\ref{lambdaK}) connects chaotic dynamics and geometry. This connection paves the way for studying thermodynamics from a geometric perspective.

Moreover, we find that the Gaussian curvature $K_{HL}$ of the unstable null circular orbits in Fig.~\ref{HLKT} is consistently negative, in full agreement with the result of Ref.~\cite{GC01}. In Ref.~\cite{GC01}, the authors adopted an alternative approach based on the Hadamard theorem, demonstrating that $K<0$ corresponds to unstable circular orbits, while $K>0$ indicates stable ones. Our results show that the Gaussian curvature on the light ring is strictly negative, supporting their conclusion. Additionally, after the black hole undergoes a first-order phase transition, specifically on the right side of the spinodal region, $|K|$ decreases monotonically with increasing temperature $\tilde{T}$. This behavior suggests that $K$ may be related to the order parameter, a finding consistent with other recent studies Refs.~\cite{JHEP2022,order,order01,order02}. These results indicate that the Gaussian curvature becomes multivalued whenever the black hole experiences a phase transition. 

\begin{figure*}
	\begin{minipage}{1\hsize}
		\begin{center}
			
			\subfigure[]{
				\label{F5}
				\includegraphics*[scale=0.2]{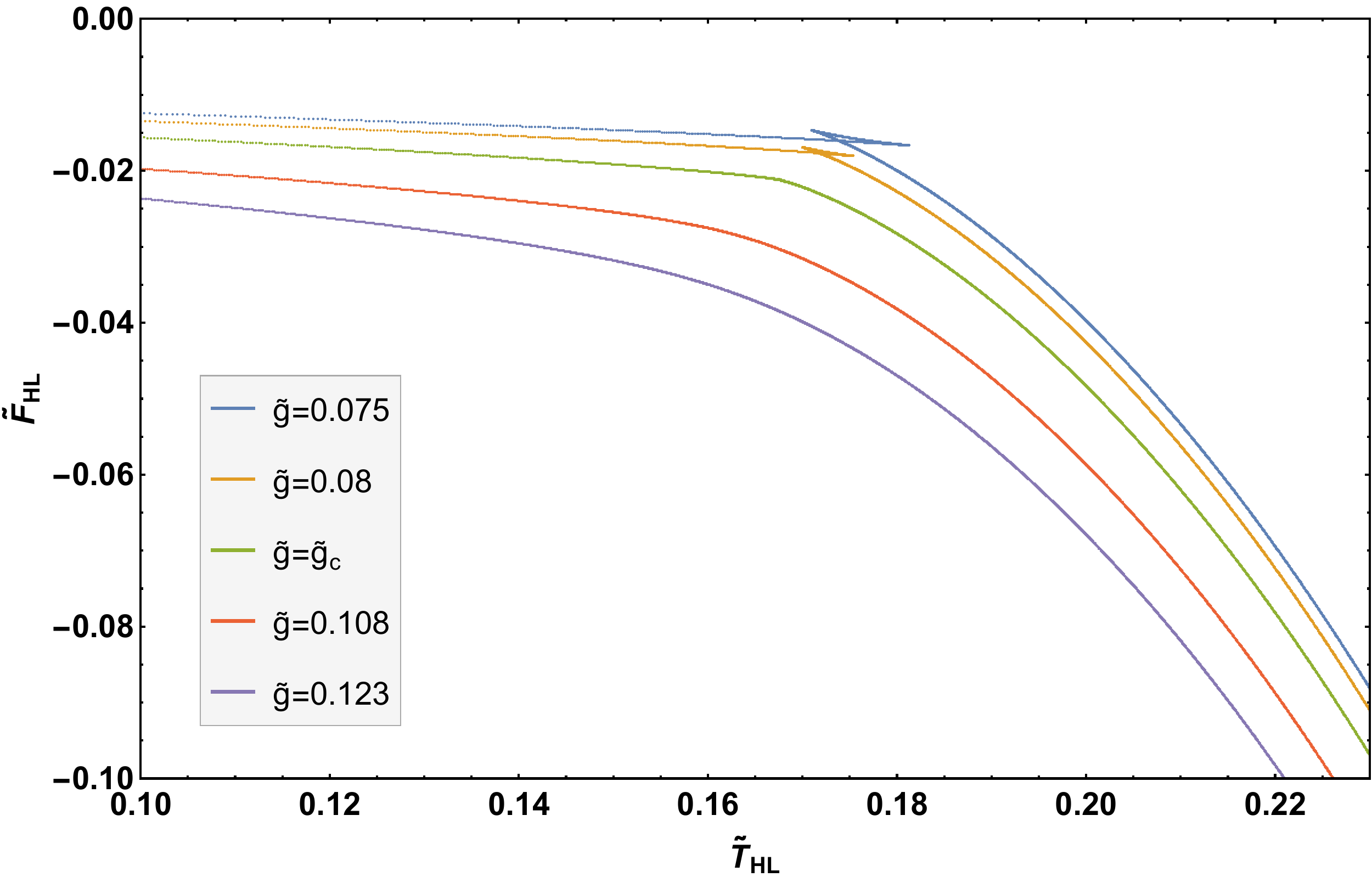}}
			\subfigure[]{
				\label{K5}
				\includegraphics*[scale=0.2]{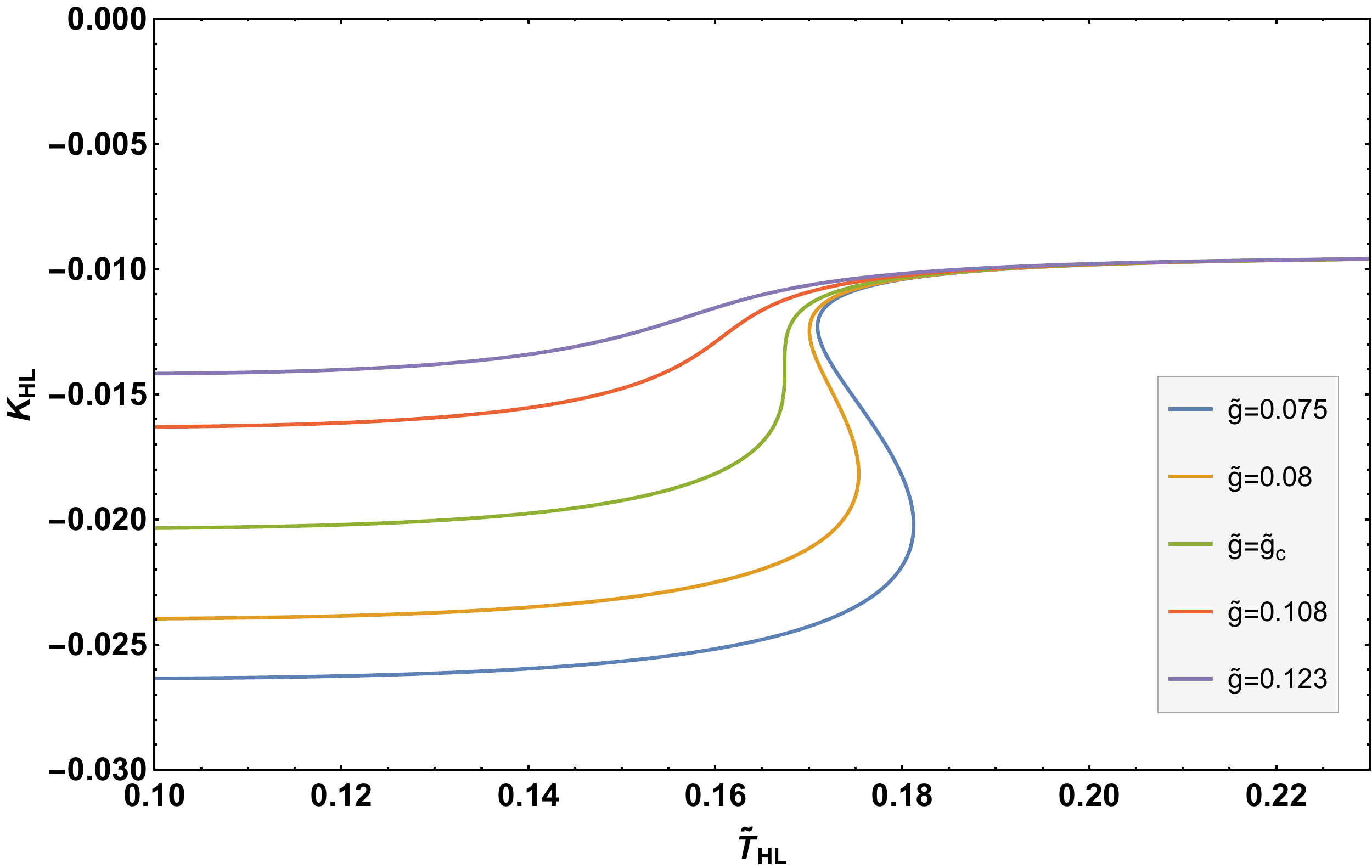}}
            
		\end{center}
		\caption{Thermodynamic and geometric signatures of first-order phase transitions in Hayward–Letelier–AdS black holes. (a) Free energy $\tilde{F}_{HL}$ and (b) Gaussian curvature $K_{HL}$ versus temperature $\tilde{T}_{HL}$ for a Hayward-Letelier-AdS black hole at various values of magnetic charge ($\tilde{g}_c=0.090,~a=0.6$). When the first‑order phase transition is absent, the Gaussian curvature and the free energy simultaneously return to monotonic behavior.
        }
		\label{Fig.4}
	\end{minipage}
\end{figure*}

In contrast, as illustrated in Fig.~\ref{Fig.4}, for the case $\tilde{g}>\tilde{g}_c$ where no first-order phase transition occurs, Fig.~\ref{Fig.4} shows that $K_{HL}(\tilde{T}_{HL})$ becomes a monotonic function of $\tilde{T}_{HL}$. Thus, the multivalued behavior of $K$ and $\lambda$ serves as a robust geometric characterization of the first-order phase transition phenomenon and its associated spinodal region, revealing a profound link between spacetime geometry and thermodynamics.

\section{Scaling exponent from Lyapunov Exponents and Gaussian Curvature}
\label{chap:4}
Scaling exponents characterize the power-law behavior of physical quantities as the system approaches the critical point. They provide important quantitative information for understanding phase transitions. At the phase transition point, the small and large black hole branches coexist, leading to discontinuities in both the null Lyapunov exponent and the Gaussian curvature. This section will examine the continuity of the null Lyapunov exponent and Gaussian curvature for the Hayward–Letelier–AdS black hole at the phase transition point $T_p$, as well as their discontinuity near $T_p$, investigating whether they can serve as order parameters and calculating their scaling exponents.

We define the null Lyapunov exponents for the small and large black hole branches within the spinodal region as $\lambda_S$ and $\lambda_L$, respectively, and the corresponding Gaussian curvatures as $K_S$ and $K_L$. Their differences are defined as $\Delta \lambda$ and $\Delta K$. When the critical point $T_p=T_c$, we have $\Delta \lambda=\Delta K=0$. We will fit the numerical coefficients of $\Delta \lambda$ and $\Delta K$ according to the following form \cite{PS01}
\begin{align}
\frac{\Delta\lambda}{\lambda_c}, \frac{\Delta K}{K_c} \sim \alpha(t-1)^\beta.
\end{align}
Here, $t=T_p/T_c$, $\alpha$ and $\beta$ are real numbers, the subscript ``$c$'' denotes the value of the Lyapunov exponent and Gaussian curvature at the critical point.

\begin{figure*}
	\begin{minipage}{1\hsize}
		\begin{center}
			
			\subfigure[]{
				\label{DL}
				\includegraphics*[scale=0.24]{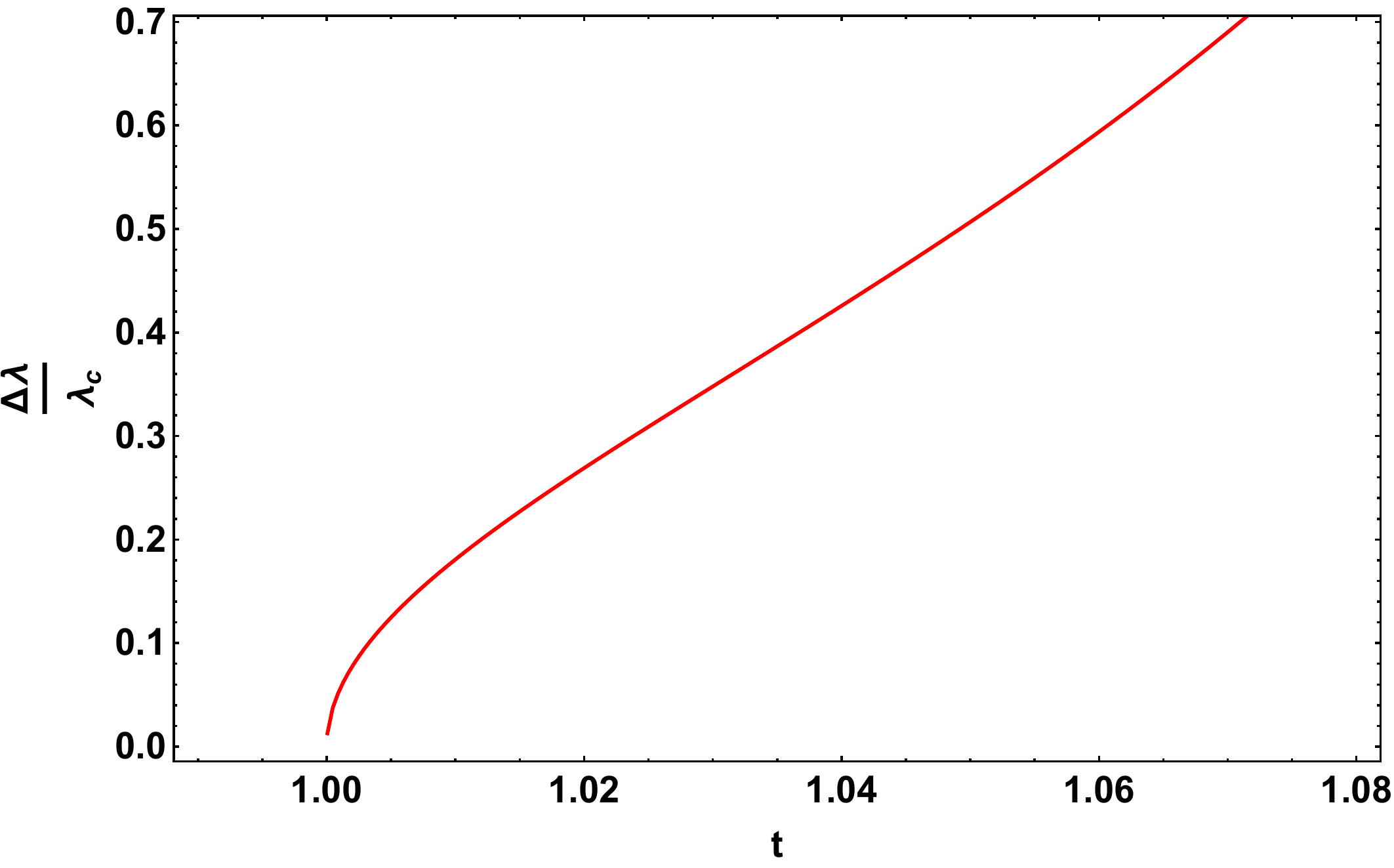}}
			\subfigure[]{
				\label{DK}
				\includegraphics*[scale=0.24]{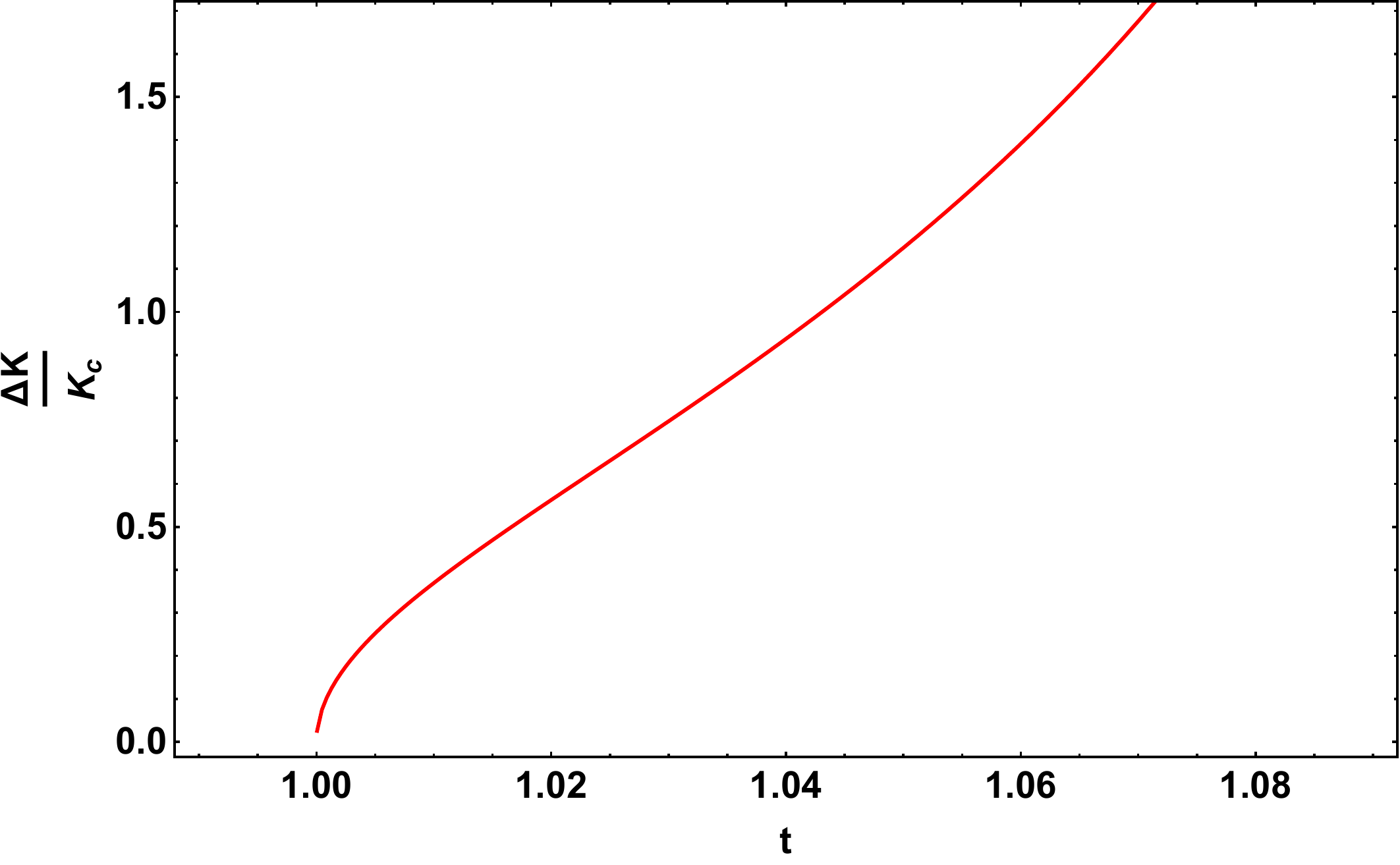}}
            
		\end{center}
		\caption{The reduced null Lyapunov exponents and Gaussian curvatures for the coexistence small and large black holes. (a) Null Lyapunov exponents and (b) Gaussian curvature $K_{HL}$ versus reduced temperature $t$ for a Hayward-Letelier-AdS black hole.
        }
		\label{Fig.5}
	\end{minipage}
\end{figure*}

As shown in Fig.~\ref{Fig.5}, we plot the normalized differences ${\Delta\lambda}/{\lambda_c}$ and ${\Delta K}/{K_c}$ as functions of the reduced temperature $t$. When $T_p=T_c$, ${\Delta\lambda}/{\lambda_c}={\Delta K}/{K_c}=0$. When $T_p$ deviates from $T_c$, both ${\Delta\lambda}/{\lambda_c}$ and ${\Delta K}/{K_c}$ take non-zero values, indicating the emergence of discontinuity. Ultimately, we obtain the following analytical expressions
\begin{align}
\frac{\Delta\lambda}{\lambda_c} \simeq 8.484(t-1)^{0.563} ,
\end{align}
and
\begin{align}
\frac{\Delta K}{K_c} \simeq 62.419(t-1)^{0.592}.
\end{align}
These results indicate that ${\Delta\lambda}/{\lambda_c}$ and ${\Delta K}/{K_c}$ exhibit order parameter-like behavior. However, their scaling exponents deviate from the exponent of $1/2$ observed in Reissner-Nordström-AdS black holes, suggesting that the geometry and dynamics of regular black holes exhibit richer critical behavior compared to singular black holes.

\section{Discussion}\label{chap:5}

In this work, we systematically investigate the profound connections between spacetime geometry, thermodynamics, and chaos dynamics during first-order phase transitions of black holes. While the standard thermodynamic methods describe phase transitions through the behavior of thermodynamic potentials, the corresponding evolution of spacetime geometry itself demands a more direct description. Based on this consideration, we examine the possibility of employing the intrinsic geometric quantity Gaussian curvature $K$ as a new probe.

We find that during a first-order phase transition, the Gaussian curvature of unstable null orbits exhibits multivalued behavior inside the spinodal region and coincides precisely with the swallowtail structure of the free energy. This phenomenon can be understood through the known relation $K(r_{LR}) = -\lambda^2(r_{LR})$, whereby the Gaussian curvature inherits the multivalued nature of the Lyapunov exponent. However, this relation should not be interpreted as a mere translation of the dynamical results into geometric language. The key advance of our work lies in establishing a self-contained geometric diagnostic framework that operates independently of any dynamical calculation. The procedure is purely geometric: first, determining the light ring radius from the geodesic curvature of the null circular orbit in the optical metric, then computing the corresponding Gaussian curvature via the first fundamental form, and finally, checking for multivalued behavior. This self-consistent procedure yields results in precise agreement with thermodynamic predictions.

Through numerical analysis of the Hayward-Letelier-AdS black hole, we find that when the parameter $\tilde{g}$ enters the spinodal region, the free energy exhibits a swallowtail structure, while both the Gaussian curvature $K$ and the Lyapunov exponent $\lambda$ exhibit multivalued behavior. The temperature intervals over which this occurs align precisely with the thermodynamic spinodal region. Furthermore, we demonstrate that in the absence of a phase transition ($\tilde{g}>\tilde{g}_c$), the Gaussian curvature $K$ varies monotonically and the geometric multivaluedness disappears. This confirms that the multivalued behavior of the Gaussian curvature is a geometric signature of the phase transition.

Near the critical point, the scaling laws of the Gaussian curvature and the null Lyapunov exponent for the Hayward-Letelier-AdS black hole deviate from those observed in most black hole solutions, suggesting that regular black holes may exhibit more interesting dynamical and geometric behaviors.

These results advance the paradigm from standard thermodynamic or dynamical frameworks toward a purely geometric perspective. They indicate that black hole first-order phase transitions can be fundamentally understood as a branching behavior in the curvature structure of spacetime. Our work provides the direct evidence of the accompanying degeneracy in spacetime geometry during such phase transitions and demonstrates that Gaussian curvature can serve as a self-consistent diagnostic probe. This opens a new
pathway for studying black hole phase structures from a geometric perspective.

\section*{Appendix A: Self-Consistent Free Energy}
Ref.~\cite{HLT01} noted that for the Hayward–Letelier–AdS black hole, the temperature defined by surface gravity, $T_k$, disagrees with the temperature obtained from the first law, $T_H$, so the first law must be modified. This modification introduces a correction factor $W=r_+^3/(r_+^3+g^3)$ such that
\begin{align}
T_k = WT_H.
\end{align}
Consequently, in the canonical ensemble, the ADM mass $M$ is replaced by a corrected mass $\mathcal{M}$ that satisfies
\begin{align}
d\mathcal{M} = WdM,
\end{align}
whose explicit form can be obtained by numerical integration of this relation. The resulting self‑consistent free energy is therefore
\begin{align}
F = \mathcal{M}-T_kS.\label{FC}
\end{align}
Therefore, a self‑consistent thermodynamic analysis should be performed by plotting the corrected free energy Eq.~(\ref{FC}) against $T_k$.

Notably, probes such as the Gaussian curvature and the Lyapunov exponent do not depend on the corrected mass $\mathcal{M}$. This offers considerable convenience for such phase transition analyses in black holes with intricate thermodynamic properties.

\section*{Appendix B: Second-Order Phase Transitions}
As a supplementary discussion of the phase transition behavior, we examine whether the Gaussian curvature can also serve as an effective probe for second-order phase transitions. At the critical point $T_{p2}$, the thermodynamic conditions
\begin{align}
\frac{\partial T}{\partial r_+} = \frac{\partial^2 T}{\partial r_+^2} = 0
\end{align}
are satisfied, signaling a second-order phase transition \cite{Kubiznak:2016qmn}.

\begin{figure*}
	\begin{minipage}{1\hsize}
		\begin{center}
			
			\subfigure[]{
				\label{dK}
				\includegraphics*[scale=0.24]{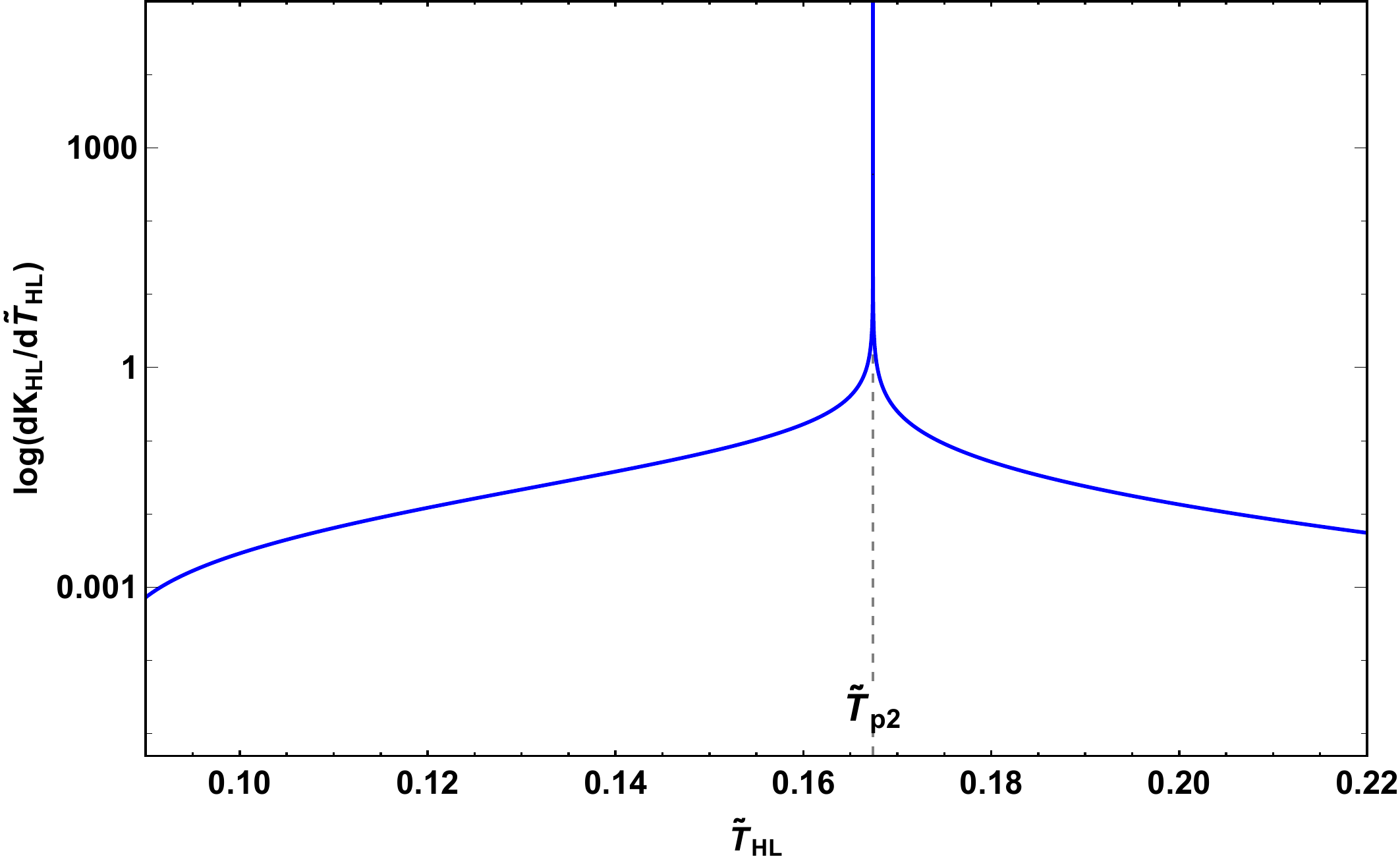}}
			\subfigure[]{
				\label{dL}
				\includegraphics*[scale=0.24]{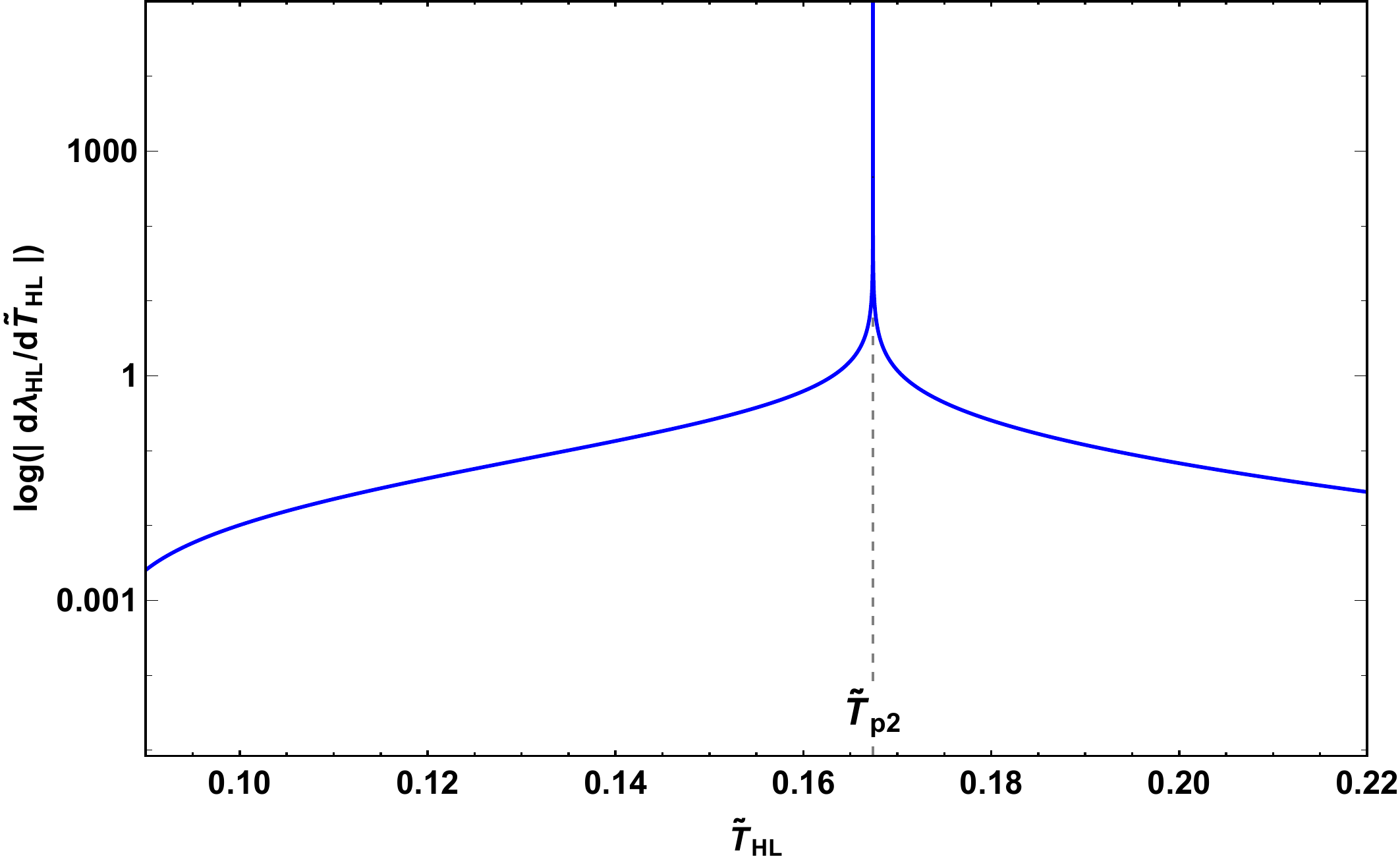}}
            
		\end{center}
		\caption{$\log\left(\frac{dK_{\mathrm{HL}}}{d\tilde{T}_{\mathrm{HL}}}\right)$ and $\log\left(\left|\frac{d\lambda_{\mathrm{HL}}}{d\tilde{T}}\right|\right)$ at the critical point versus reduced temperature for a Hayward-Letelier-AdS black hole.
        }
		\label{Fig.6}
	\end{minipage}
\end{figure*}

As shown in Fig.~\ref{Fig.6}, when the second-order phase transition occurs, the derivatives (slopes) of both the Gaussian curvature $K$ and the Lyapunov exponent $\lambda$ with respect to temperature diverge at $T_{p2}$. This behavior can be understood by following the continuous evolution of the system: as the black hole parameter (the magnetic charge $g$) is tuned from the first-order phase transition regime toward the critical point, the multivalued structure of $K(T)$ progressively shrinks. At exactly the critical point, the three branches of the swallowtail merge, and the system finds itself precisely at the threshold between multivalued and single-valued behavior. During this shrinking process, the slope of the intermediate branch of $K(T)$ increases, eventually diverging at the critical point. This divergent behavior is directly analogous to the well-known divergence of the heat capacity at a critical point in standard thermodynamics \cite{Kubiznak:2016qmn}. Such an analysis procedure also applies to the divergent behavior of the derivative of the Lyapunov exponent.

This establishes that the Gaussian curvature serves as an effective probe for phase transitions: its multivaluedness signals a first-order phase transition, while the divergence of its derivative identifies a second-order phase transition.

\section*{Appendix C: Zeroth-Order Phase Transitions}
In this paper, we have mainly investigated the effectiveness of Gaussian curvature as a probe for first- and second-order phase transitions of black holes. However, as a natural extension, here we briefly discuss its potential extension to zeroth-order phase transitions.

A zeroth-order phase transition is characterized by a discontinuity in the physically realized free energy under the condition of global free energy minimization. Unlike a first-order phase transition, a zeroth-order phase transition involves a direct jump between two distinct branches of the free energy at the phase transition point. 

From the geometric framework perspective, the discontinuity in the black hole solution branches implies that geometric quantities dependent on the spacetime properties of these branches can similarly exhibit discontinuities. For instance, the Gaussian curvature may also jump between different branches. This is because different solution branches generically possess distinct metric functions, and hence different values of $K$ at the same temperature.

We note, however, that the Hayward-Letelier-AdS black hole studied in this work belongs to a class of models whose thermodynamic phase structure is dominated by first- and second-order phase transitions. Investigating the possible existence of zeroth-order phase transitions in this or related regular black hole models would constitute an interesting study in its own right, and we leave this as a promising direction for future work. The geometric framework established here provides a solid theoretical foundation for such an extension.

\section{Acknowledgements}
This work was supported by the National Natural Science Foundation of China (Grant Nos. 12473001, 12575049, and 12533001), the National SKA Program of China (Grant Nos. 2022SKA0110200 and 2022SKA0110203), the China Manned Space Program (Grant No. CMS-CSST-2025-A02), and the 111 Project (Grant No. B16009).

\bibliography{reference}

\end{document}